\documentstyle[12pt]{article}
\newcommand{\be}{\begin{equation}}
\newcommand{\inn}{\!\cdot\!}
\newcommand{\veps}{\varepsilon}
\newcommand{\eg}{{\it e.g.,}\ }
\newcommand{\ie}{{\it i.e.,}\ }
\newcommand{\tr}{{\rm Tr}}
\newcommand{\bea}{\begin{eqnarray}}
\newcommand{\eea}{\end{eqnarray}}
\newcommand{\ba}{\begin{array}}
\newcommand{\ea}{\end{array}}
\newcommand{\ee}{\end{equation}}

\expandafter\ifx\csname mathbbm\endcsname\relax

\else

\fi
\begin{document}
\begin{titlepage}
\hfill
\vbox{
    \halign{#\hfil         \cr
           IPM/P-2002/001 \cr
           hep-th/0201249  \cr
           } % end of \halign
      }  % end of \vbox
\vspace*{20mm}
\begin{center}
{\Large {\bf Gauge Invariant Operators and Closed String Scattering
in Open String  Field Theory}\\ }

\vspace*{15mm}
\vspace*{1mm}
{Mohsen Alishahiha$^{a,}$\footnote{alishah@theory.ipm.ac.ir}  and  Mohammad  
R. Garousi$^{b,a,c,}$\footnote{garousi@theory.ipm.ac.ir}}\\

\vspace*{1cm}

{\it$^a$ Institute for Studies in Theoretical Physics and Mathematics (IPM)\\
P.O. Box 19395-5531, Tehran, Iran\\
$^b$ Department of Physics, Ferdowsi University,
Mashhad, Iran\\
$^c$ Department of Physics, University of Birjand,
Birjand, Iran}

\vspace*{1cm}
%%\maketitle
\end{center}

\begin{abstract}
Using the recent proposal for the observables in open string field theory,  
we explicitly
compute the coupling of  closed string tachyon and massless states with the 
open string
states up to level two. Using these couplings,  we then calculate the tree 
level S-matrix
elements of  two closed string tachyons or two massless states in the open 
string field theory. Up to some contact terms, the results  reproduce 
exactly the corresponding 
amplitudes in the bosonic string theory.
\end{abstract}
%\vskip 6cm

\end{titlepage}
\newpage
\section{Introduction}
The open string tachyon condensation has attracted much interest recently.  
Regarding the
recent development in string field theory, (for example see \cite{KO} and 
\cite{ABGKM} and their references)
it is believed that the open string field theory \cite{WITTEN} might provide a
direct approach to study the physics of string theory tachyon and could give
striking evidence for the tachyon condensation conjecture regarding the 
decay of unstable
D-branes or the annihilation of brane anti-brane system \cite{SEN}.  
Therefore it would be
very interesting to study and develop the structure of the open string 
field theory itself.

On the other hand the most difficult part of the Sen's conjecture for open 
string
tachyon is the way the closed string emerges in the tachyonic vacuum . 
So it would be a natural question to ask that how one can see the closed 
string states 
in the open string field theory. In fact it has been
shown that the off-shell closed strings arise because certain one-loop 
open string
diagrams can be cut in a manner that produces a closed string pole \cite{FGST}. 
Therefore unitarity implies that they should also appear as asymptotic states. 
Of course one can not remedy this by adding an explicit closed-string 
field to the theory. This would just double the residue of the pole. 
They can not be also considered as a bound states, since they appear 
in the perturbation theory.
Closed string in open string field theory has been studied in several 
papers including
\cite{{TU},{ST1}, {STO}, {ZW}}.

In an other attempt but related to the closed string states in the
open string field theory, the gauge invariant operators in open
string field theory have been considered in \cite{{HN}, {GRSW}}.
These gauge invariant operators could also provide us the on-shell
closed string in the open string field theory. In fact these
operators are parameterized by on-shell closed string vertex
operators and can arise from an open/closed transition vertex that
emerged in one-loop open string theory. Actually this open/closed
vertex was studied in \cite{ZW} where it was shown that
supplemented  with open string vertex it would generate a cover of
the moduli spaces of surfaces involving open and closed string
punctures.

It has also been suggested in \cite{{HN},{GRSW}} that the  correlation 
function of these gauge invariant operators could be interpreted as the 
on-shell scattering amplitude of the closed strings from D-brane.
This is the aim of this article to study this correspondence in more detail.
We shall study the scattering amplitude of two closed string states off a 
D-brane in the framework of the open string field theory by making use 
of these gauge invariant operators.

The paper is organized as follows.  In section 2, we shall review
the open string field theory action as well as the gauge invariant
operators introduced in \cite{{HN},{GRSW}}. In section 3 we will
evaluate the scattering amplitude of the closed strings  in the
framework of string field theory. In section 4 the
same scattering amplitudes will be obtained in the bosonic string theory
where we will show that up to some contact terms, the results are
in agreement with the open string field theory results. The
section 5 is devoted to the discussion and some comments.

\section{ Open String Field Theory}

In this section we shall review the open string field theory and the
structure of the gauge invariant operators which could provide
observables of the open string field theory.

\subsection{Cubic string field theory action}

The cubic open string field theory action is given by
\bea
S(\Psi)&=&-\frac{1}{2\alpha'}\left(\int \Psi * Q\Psi+{2g_o\over 3} 
\Psi*\Psi*\Psi\right)\;\; ,
\label{SA}
\eea
which is invariant under the gauge transformation 
$\delta\Psi=Q\Lambda+g_o\Psi * \Lambda-g_o\Lambda * \Psi$.
Here $g_o$ is the open string coupling, $Q$ is the BRST charge and the 
string field, $\Psi$, is a ghost number
one state in the Hilbert space of the first-quantized
string theory which can be expanded using the Fock space basis as
\footnote{Here, we use the convention fixed in \cite{GM} that uses the V 
and N matrices for projecting a space-time field 
to its component in the world-volume and transverse spaces, respectively.
So in this convention $\mu,\nu=0,1,2,...,25$, and $A_{\mu}\alpha_{-1}^{\mu}=
A\inn V\inn\alpha_{-1}+A\inn N\inn\alpha_{-1}$.
Our conventions also set $\alpha'=2$.}
\bea
|\Psi\rangle &=&\int d^{p+1}k\;(\phi+A_{\mu}\alpha_{-1}^{\mu}+i\alpha
b_{-1}c_0+{i\over \sqrt{2}}B_{\mu}\alpha_{-2}^{\mu}+{1\over \sqrt{2}}
B_{\mu\nu} \alpha_{-1}^{\mu}\alpha_{-1}^{\nu}  \cr
&+&  \beta_0b_{-2}c_0+\beta_{1}b_{-1}c_{-1}+ik_{\mu}
\alpha_{-1}^{\mu}b_1c_0
+\cdots ) c_1 |k\rangle\;\; . \nonumber
\eea
The gauge invariance of (\ref{SA}) can be fixed can  
by choosing Feynman-Siegel gauge $b_0 |\Psi\rangle=0$.
In this gauge the truncated field up to level two reads
\bea
|\Psi\rangle &=&\int d^{p+1}k \left(\phi(k)+A_{\mu}(k)\alpha_{-1}^{\mu}
+{i\over \sqrt{2}}B_{\mu}(k)\alpha_{-2}^{\mu}+\right.\cr
&+&\left. {1\over \sqrt{2}}B_{\mu\nu}(k) \alpha_{-1}^{\mu}\alpha_{-1}^{\nu}
+\beta_{1}(k)b_{-1}c_{-1}\right) c_1 |k\rangle\;\; . \nonumber
\eea
The corresponding string vertex is given by
\bea
\Psi(0) &=&\int d^{p+1}k\left[\phi(k)c(0)+iA_{\mu}(k)c\partial X^{\mu}(0)+
\frac{i}{\sqrt{2}}B_{\mu}(k)c\partial^2 X^{\mu}(0)
\right.\cr 
&-&\left.\frac{1}{\sqrt{2}}B_{\mu\nu}(k)c\partial X^{\mu}\partial X^{\nu}(0)-
{1\over 2}\beta_{1}(k)\partial^2 c(0)
\right] e^{2ik\inn X(0)}\;\; .
\label{STRING}
\eea
In writing the above vertex, we have used the doubling trick \cite{GM}. 
Hence,  the world-sheet field $X^{\mu}(z)$ in above equation is only 
holomorphic part of $X^{\mu}(z,\bar{z})$.

To make sense out of the abstract form of the open string field
theory action, one can use CFT method. In this method we usually
use the conformal mapping and  calculation of the correlation
function of a CFT on a disk or upper-half plane \cite{{LPP},
{RZ}}.  In the CFT language the $n$-string vertex is defined by \footnote{
We assume that there is a normal order sign between fields at different
points in the correlation functions.}
\bea 
&\int \Psi*\Psi*\cdots *\Psi=\left\langle\; f^{(n)}_1\circ\Psi(0)\;
f^{(n)}_2\circ\Psi(0)\cdots f^{(n)}_n \circ\Psi(0)\;
\right\rangle_{UHP}\;\; , & \nonumber
\eea 
where $f^{(n)}_k\circ\Psi(0)$ denotes the conformal transformation
of the vertex operator $\Psi(0)$ by the conformal map $f^{(n)}_k$.
Here $\langle\;\; \rangle_{UHP}$ denotes correlation function on the
upper-half plane and the conformal map $f^{(n)}_k$  is defined as
\bea 
f^{(n)}_k(z_k)=g\left(e^{{2\pi i\over
n}(k-1)}\;\left(\frac{1+iz_k}{1-iz_k}\right)^{2\over n}\right)\; , &&
\;\;\;\;\;\; 1\leq k\leq n\;\; , \nonumber 
\eea 
where $g(\zeta)=-i\frac{\zeta-1}{\zeta+1}$. Therefore the open string
action can be calculated as following in terms of correlation
functions of the CFT on the UHP 
\bea
S&\!=\!&-\frac{1}{4}\left\langle\; f^{(2)}_2\circ\Psi(0)
f^{(2)}_1\circ(Q\Psi(0))+{2g_o\over 3} \; f^{(3)}_1\circ\Psi(0)\;
f^{(3)}_2\circ\Psi(0) f^{(3)}_3 \circ\Psi(0)\;\right\rangle_{UHP}.
\nonumber
\eea 
Form this expression the kinetic terms up to level two fields read 
\bea 
S_{\rm quad}&=&\int
d^{p+1}x(-\frac{1}{2}\partial_{\mu}\phi\partial^{\mu}
\phi+\frac{1}{4}\phi^2-\frac{1}{2}\partial_{\mu}A_{\nu}\partial^{\mu}
A^{\nu}- \frac{1}{2}\partial_{\mu}B_{\nu}\partial^{\mu}B^{\nu}-
\frac{1}{4}B_{\mu} B^{\mu}\cr
&\!-\!&\frac{1}{2}\partial_{\lambda}B_{\mu\nu}\partial^{\lambda}
B^{\mu\nu}
-\frac{1}{4}B_{\mu\nu}B^{\mu\nu}+\frac{1}{2}\partial_{\mu}\beta_1
\partial^{\mu}\beta_1+\frac{1}{4}\beta_1^2)\;\; ,
\label{squad}
\eea
which can be used to write the space-time propagators of the corresponding 
fields.

\subsection{ Gauge invariant operator}

The gauge invariant operators in string field theory have been 
constructed in \cite{HN, GRSW}. The general form of these operators 
are given by ${\cal O}=g_c\int V \Psi$,
where $g_c$ is the closed string coupling and $V$ is an on-shell closed 
string vertex operator with dimension (0,0).
In order to be gauge invariant, the closed string vertex operator has to be
inserted at the midpoint of open string. From open string point of view, 
$ V$ is
an operator which acts on a string field. Given any on-shell closed 
string vertex
operator $V$, the gauge invariant operator ${\cal O}$ can be obtained, using
the CFT method, in terms of the open string field 
\be
{\cal O}=g_c\int V\Psi=g_c\left\langle\;c{\cal V}(i)\; 
{\bar c}{\bar {\cal V}}(-i)f^{(1)}_1\circ\Psi(0)\;\right\rangle_{UHP}\;\; ,
\label{GO1}
\ee
where $f^{(1)}_1=\frac{2z}{1-z^2}$ and 
${\cal V}(z){\bar {\cal V}}({\bar z})$ 
is the matter part of the closed string vertex operator.

This form of the gauge invariant operator can be understood from the 
closed/open vertex studied in \cite{ZW}, where is was shown that the extended 
open string field theory with the action 
\be
S=-\frac{1}{4}\left(\int \Psi * Q\Psi +{2g_o\over 3}\Psi *\Psi *\Psi\right)
+g_c\int V\Psi\;\; ,
\label{ACP}
\ee
with $V$ being an on-shell closed string vertex defined at the midpoint 
of the open string, would provide a theory which covers the full moduli 
space of the scattering amplitudes of open and closed string with a 
boundary. We note, however, that scattering amplitudes of  open and 
closed string
with a boundary are actually the closed string scattering off a D-brane.  
We should then be able to reproduce the closed string scattering 
amplitudes in the framework of the open string field theory. 
In the next section we are going to write down the explicit form of the
gauge invariant operator as well as their correlation function among 
themselves to see to what extent we can reproduce the known results of
the closed string scattering amplitudes from a D-brane
in the bosonic string theory \cite{{KT},{CLR}}.

\section{Scattering amplitudes in  string field theory}

In this section we will consider the gauge invariant operators in
the string field theory. Using CFT method we shall compute the
explicit form of the operators in terms of space-time open string
fields. According to the proposed action
(\ref{ACP}) the result can be thought as an space-time action
representing the closed/open vertex. We shall also compute the
correlation function of these operators among themselves. These
correlators should be interpreted as  the closed string
scattering amplitude off a D-brane. 
We shall perform all of our computations in the
level truncation up to level two. In evaluating the correlations
(\ref{GO1}), one needs the transformation of the vertex
(\ref{STRING}) under conformal map $f_1^{(1)}$. Using the
following propagator 
\be \langle \; X^{\mu}(z) X^{\nu}(w)\;
\rangle=-\eta^{\mu\nu}\ln(z-w)\;\; ,
\label{pro} 
\ee 
one finds that the different terms in (\ref{STRING}) transform 
under a general conformal map $f$ as 
\bea 
f\circ(ce^{2ik \inn  X})&=&{f'}^{2
k^2-1}ce^{2ik\inn  X},\cr
f\circ(c\partial X^{\mu}e^{2ik\inn
X})&=&{f'}^{2 k^2} (\partial X^{\mu}- ik^{\mu}f''/{f'}^2 )ce^{2ik\inn  X}, \cr
f\circ(c\partial^2X^{\mu}e^{2ik\inn  X})&=&{f'}^{2 k^2+1}[
\partial^2X^{\mu}
 +(f''/{f'}^2)\partial X^{\mu}\cr 
&-&ik^{\mu}/6\;(4f'''/{f'}^3-3
(f''/{f'}^2)^2)]ce^{2ik\inn  X}, \label{CONT}\\
f\circ(c\partial X^{\mu}\partial X^{\nu}e^{2ik\inn  X})&=&{f'}^{2 k^2+1}
[\partial X^{\mu}\partial X^{\nu}-2i(f''/{f'}^2)k^{\{\mu}
\partial X^{\nu\}}\cr
&-&(f''/{f'}^2)^2k^{\mu}k^{\nu}-
1/12\;(2f'''/{f'}^3-3
(f''/{f'}^2)^2)\eta^{\mu\nu}]ce^{2ik\inn  X}, \cr
f\circ(\partial^2ce^{2ik\inn X})&=&{f'}^{2 k^2+1}[
\partial^2c-(f''/{f'}^2)\partial c
-(f'''/{f'}^3-2(f''/{f'}^2)^2)c]e^{2ik\inn  X}.\nonumber
\eea
Note that in the above equations the world-sheet fields on the right 
hand side are functions of $f(z)$, also $k^2=k\inn  k$.
The function $f_1^{(1)}$ and its derivatives at point $z=0$ that should 
be inserted in  the above transformations are: 
$f_1^{(1)}(0)=0,\;{f_1'}^{(1)}(0)=2,\; {f_1''}^{(1)}(0)=0,\;
{f_1'''}^{(1)}(0)=12$. The correlation functions over the ghost field 
that left over are $\langle c(i){\bar{c}}(-i)c(0)\rangle=2i$ and 
$\langle c(i){\bar{c}}(-i)\partial^2 c(0)\rangle=4i$.

Plugging the conformal transformation (\ref{CONT}) into the equation 
(\ref{GO1}) and  using above correlators for ghost part, we get
\bea
{\cal O}&=&2ig_c\int d^{p+1}k\;2^{2k^2}\left[\left(\frac{1}{2}\phi+
\sqrt{2}B\inn V\inn k+\frac{1}{2\sqrt{2}}B_{\mu}{}^{\mu}-\frac{1}{2}
\beta_1\right)\left\langle
{\cal V}(i){\bar {\cal V}}(-i) e^{2ik\inn  X}(0)
\right\rangle\right.\cr
&+&iA_{\mu}\;\left\langle{\cal V}(i){\bar {\cal V}}(-i)\partial 
X^{\mu}e^{2ik\inn  X}(0)\right\rangle+i\sqrt{2}B_{\mu}\;
\left\langle{\cal V}(i){\bar {\cal V}}(-i)\partial^2 X^{\mu}
e^{2ik\inn  X}(0)\right\rangle\cr
&-&\left.\sqrt{2}B_{\mu\nu}\;\left\langle{\cal V}(i){\bar {\cal
V}}(-i)\partial X^{\mu}\partial X^{\nu}e^{2ik\inn
X}(0)\right\rangle\frac{}{}\right]\;\; ,
\label{geno} 
\eea 
where all correlations should be  evaluated on the upper-half plane. 
Now we have all ingredients we need to compute the open string field theory
observable (\ref{GO1}). We will do this for both closed string
tachyon and massless states.

\subsection{Tachyon amplitude}

The matter part of the vertex  operator of  closed string tachyon inserted at
the midpoint of open string with momentum $p^{\mu}$ ($p\inn p=2$) is given by
\bea
&{\cal V}(i){\bar {\cal V}}(-i)=e^{ip.X(i)}e^{ip.D.X(-i)}\;\; , & \nonumber
\eea
where $2V=\eta+D$, and we have used the doubling trick\cite{GM}. Plugging 
this operator into equation (\ref{geno}) and using the standard propagator 
(\ref{pro}), one can evaluate the correlators in (\ref{geno}). The result is
\bea
{\cal O}(p^{\mu})&=&\frac{ig_c}{8}(2\pi)^{p+1}e^{4{\rm ln}2\;p\inn 
V\inn p}(\frac{}{}\phi+4i\;p\inn N\inn A+2\sqrt{2}\;p\inn V\inn 
B\cr
&&-8\sqrt{2}\;p\inn N\inn B\inn N\inn p+\frac{1}{\sqrt{2}}
B_{\mu}{}^{\mu}-\frac{}{}\beta_1)\; ,\nonumber
\eea
here the space-time fields are functions of $-p\inn V$.
Fourier-transforming to the position space, \eg 
$\phi(k)=\int\frac{d^{p+1}x}{(2\pi)^{p+1}}\; 
\phi(x)\;e^{-ik.x}$,
the operator ${\cal O}$ becomes
\bea
{\cal O}(p^{\mu})&=&{ig_c\over 8}\int d^{p+1}x\; e^{ip.x}\;
\left({\tilde \phi}(x)+ 4i\;p\inn N\inn{\tilde A}(x) +
2\sqrt{2}\;p\inn V\inn {\tilde B}(x)\right.\cr
&&\left.-8\sqrt{2}\;p\inn N \inn{\tilde B}(x)\inn N\inn p+
\frac{1}{\sqrt{2}}{\tilde B}_{\mu}{}^{\mu}(x)-{\tilde \beta_1}(x)\right)\;\; ,
\label{INAC}
\eea
where the tilde sign over fields means, {\it e.g.}
${\tilde \phi}(x)=e^{-4{\rm ln}2\; \partial^2}\phi(x)$.
According to the proposed action (\ref{ACP}) the expression (\ref{INAC})
is space time action representing the coupling of the closed string
tachyon with the open string fields.

Having a gauge invariant operator one would proceed to compute the correlation
function of this gauge invariant operator. Since this operator is supposed to
be state corresponding to the on-shell closed string, therefore the correlation
function of this operator should  give the scattering amplitude of the closed
string off a D-brane. Now we are going to compute explicitly this correlation
function to see if we can reproduce the corresponding amplitudes in the
bosonic string theory. We shall do this up to level two truncation in
open string field.

Consider the following two point function
\be
\langle\; {\cal O}(p_1){\cal O}(p_2)\; \rangle\;\; ,
\label{corr}
\ee
where $\langle\;\;\rangle$ denotes the correlation function in the
string field theory. According to what we have said, this should
be interpreted as the $S$-matrix elements of two closed string
states. In order to evaluate the above correlation, one needs the
propagator of the space time open string fields which can be
obtained from the kinetic term of the string field theory action
in (\ref{squad}), \ie 
\bea
\langle \phi(x)\phi(y)\rangle&=&-i\int d^{p+1}k\frac{e^{ik\inn (x-y)}}
{k^2-\frac{1}{2}},\;\;
\langle A^{\mu}(x) A^{\nu}(y)\rangle=-i\int d^{p+1}k
\frac{\eta^{\mu\nu}e^{ik\inn (x-y)}}{k^2},\cr
\langle \beta_1(x)\beta_1(y)\rangle&=&i\int d^{p+1}k
\frac{e^{ik\inn (x-y)}}{k^2+\frac{1}{2}},\;\;\;\;\,
\langle B^{\mu}(x) B^{\nu}(y)\rangle=-i\int d^{p+1}k
\frac{\eta^{\mu\nu}e^{ik\inn (x-y)}}{k^2+\frac{1}{2}},\cr 
\langle B^{\mu\nu}(x) B^{\lambda\rho}(y)\rangle&=&-\frac{i}{2}\int 
d^{p+1}k\frac{(\eta^{\mu\lambda}\eta^{\nu\rho}+\eta^{\mu\rho}
\eta^{\nu\lambda})e^{ik\inn (x-y)}}{k^2+\frac{1}{2}}\;\;.
\label{propag}
\eea
Now inserting (\ref{INAC}) into (\ref{corr}) and using above propagators, 
one finds
\bea
\langle\; {\cal O}(p_1){\cal O}(p_2)\; \rangle &=&
\frac{ig_c^2}{2}(2\pi)^{p+1}\delta^{p+1}(p_1+p_2)
\{\frac{e^{8(s-\frac{1}{2}){\rm ln}2}}{2s-1}-
\frac{e^{8s{\rm ln}2}p_1\inn N\inn p_2}{2s}\cr
&&+\frac{e^{8(s+\frac{1}{2}){\rm
ln}2}[\frac{1}{2}(p_1\inn N\inn
p_2)^2+
\frac{1}{32}(s+\frac{1}{2})-\frac{1}{8}]}{2s+1}
\}\;\; ,\nonumber
\eea 
where $s=p_1\inn V\inn p_1$. In the above expression, those terms in each pole
which are proportional to the denominator give contact terms in
which we are not interested. Hence, the pole structure of the 
amplitude is
\bea
\langle\; {\cal O}(p_1){\cal O}(p_2)\; 
\rangle &=&\frac{ig_c^2}{2}(2\pi)^{p+1}
\delta^{p+1}(p_1+p_2)\{\frac{1}{2s-1}-
\frac{p_1\inn N\inn p_2}{2s}\cr
&&+\frac{\frac{1}{2}(p_1\inn N\inn p_2)^2-\frac{1}{8}}{2s+1}+
\cdots\}\;\; ,
\label{cortach}
\eea
where dots represent some contact terms. We shall show that the above 
poles exactly reproduce the s-channel poles of the corresponding 
amplitude in the bosonic string theory.

\subsection{Graviton amplitude}

As an other  example, let us to consider the massless states scattering off
a D-brane in the framework of the string field theory. To do this we need to
find observables corresponding to the on-shell massless closed string states
which could be dilaton, graviton or Kalb-Ramond field.
In other words we need to compute the operator (\ref{geno}) for the 
corresponding vertex operators.
The matter part of these vertex operators inserted at the midpoint are given by
 \bea
{\cal V}(i){\bar {\cal V}}(-i)&=&(\veps \inn D)_{\mu \nu}\partial 
X^{\mu}(i)e^{ip\inn X(i)} \partial X^{\nu}(-i)
 e^{ip\inn D\inn X(-i)}\;\; ,\nonumber
\eea
with $p_{\mu}p^{\mu}=0=p^{\mu}\veps_{\mu \nu}=\veps_{\mu\nu}p^{\nu}$. For
graviton we have $ \veps_{\mu\nu}=\veps_{\nu\mu}$ and $\veps^{\mu}_{\mu}=0$ 
\cite{GM}.

Plugging above  closed string vertex operator into the equation (\ref{geno}) 
and  using the world-sheet propagator (\ref{pro}), one finds
\bea
{\cal O}(\veps,p)&=&\frac{ig_c}{8}(2\pi)^{p+1}e^{4{\rm ln}2\;p\inn V\inn p}
\left(\phi\; a+4iA_{\mu}b^{\mu}+2\sqrt{2}B_{\mu}c^{\mu}-8\sqrt{2}
B_{\mu\nu}d^{\mu\nu}-\beta_1 a\right)\;\; ,\nonumber
\eea
here the space-time fields are functions of $-p\inn V$. 
The kinematic factors $a,b^{\mu},c^{\mu}$, and $d^{\mu\nu}$ are
\bea
a&=&\tr(\veps\inn D)-p\inn D\inn\veps\inn D\inn p\;\; ,\cr
b^{\mu}&=&a\; p\inn N^{\mu}+p\inn D\inn\veps\inn D^{\mu}-\veps^{\mu}\inn 
D\inn p\;\; , \cr
c^{\mu}&=&a \;p\inn V^{\mu}-4p\inn D\inn \veps\inn D^{\mu}-4\veps^{\mu}
\inn D\inn p\;\; ,\cr
d^{\mu\nu}&=&a(p\inn N^{\mu} p\inn N^{\nu}-\frac{1}{16}\eta^{\mu\nu})+
2(\veps\inn D)^{\{\mu\nu\}}+2p\inn D\inn\veps\inn D^{\{\mu}p\inn N^{\nu\}}-
2\veps^{\{\mu}\inn D\inn p\; p\inn N^{\nu\}}.\nonumber
\eea
Fourier-transforming the open string fields to the position space, we get
\bea
{\cal O}(\veps,p)&=&\frac{ig_c}{8}\int d^{p+1}x({\tilde \phi}(x) a
+4i{\tilde A}_{\mu}(x)b^{\mu}\cr
&+&2\sqrt{2}{\tilde B}_{\mu}(x)c^{\mu}-8\sqrt{2}
{\tilde B}_{\mu\nu}(x)d^{\mu\nu}-{\tilde \beta}_1(x) a)\;\; ,\nonumber
\eea
where the tilded fields are defined the same as those in (\ref{INAC}). 
Inserting above operator in (\ref{corr}) and using the space-time 
propagators (\ref{propag}), one can evaluate the scattering amplitude 
of two massless states from D-brane, that is
\bea
\langle{\cal O}(\veps_1,p_1){\cal O}(\veps_2,p_2)\rangle &=&
\frac{ig_c^2}{2}(2\pi)^{p+1}\delta^{p+1}(p_1+p_2)
\{\frac{e^{8(s-\frac{1}{2}){\rm ln}2}a_1a_2}{2s-1}-
\frac{e^{8s{\rm ln}2}b_1\inn b_2}{2s}\cr
&&+\frac{e^{8(s+\frac{1}{2}){\rm ln}2}[\frac{1}{32}
c_1\inn c_2+\frac{1}{2}\tr(d_1\inn d_2)-\frac{1}{256}a_1a_2]}{2s+1}
\}\cr
&=&\frac{ig_c^2}{2}(2\pi)^{p+1}\delta^{p+1}(p_1+p_2)
\{\frac{a_1a_2}{2s-1}-\frac{b_1\inn b_2}{2s}\cr
&&+
\frac{\frac{1}{32}c_1\inn c_2+
\frac{1}{2}\tr(d_1\inn d_2)-\frac{1}{256}a_1a_2}{2s+1}+\cdots\}\;\; ,
\label{agrav}
\eea
where dots represent some contact terms.

So far we have obtained the $S$-matrix elements of two closed
string states in the framework of the open string field theory. 
Although we have only been able to find the $s$-channel poles using
level truncated open string field, we will see that the $t$-channel 
can be also obtained by taking into account the infinite terms coming from 
the all level in the open string field. We will back to this point later 
in the conclusion.
Now we are going to compare these results with the string
theory computation. 
\section{Scattering amplitudes in string theory}
In this section we study the scattering amplitude of two closed string 
states from D-brane in the bosonic string theory. 
\subsection{Tachyon amplitude}
Scattering amplitude of two closed string tachyons from D-brane in the 
bosonic string theory is given by the following correlation function:
\bea
A&\sim&\int d^2z_1d^2z_2\langle e^{ip_1\inn X(z_1)}
\;e^{ip_1\inn D\inn X({\bar z}_1)}\;e^{ip_2\inn X(z_2)}
\;e^{ip_2\inn D\inn X({\bar z}_2)}\rangle\;\; .
\nonumber
\eea
Using the propagator (\ref{pro}) it is straightforward calculations to 
evaluate the above correlation and show that the resulting integrand is 
$SL(2,R)$ invariant.
Fixing this symmetry by choosing $z_1=iy$ and $z_2=i$, one arrives at
\bea
A=\frac{ig_c^2}{2}(2\pi)^{p+1}\delta^{p+1}(p_1+p_2)B(-1-t/2,-1+2s)\;\; ,
\label{atach}
\eea
where $t=-(p_1+p_2)^2$ and $s=p_1\inn V\inn p_1$.

Consider the following definition of the beta function:
\bea
B(\alpha,\beta)&=&\sum_{n=0}^{\infty}\frac{1}{\alpha+n}\frac{(-1)^n}{n!}
(\beta-1)\cdots(\beta-n)\;\; .
\label{beta}
\eea
From this expression we see that the beta 
function has simple poles at $\alpha,\beta=0,-1,-2,-3,\cdots$. 
Each term of the above expansion has 
one simple pole in the $\alpha$-channel, however, the poles in the 
$\beta$-channel appear when adding all infinite poles of the $\alpha$-channel
in the above expansion.

By making use of this expression for the beta function the 
amplitude (\ref{atach}) reads
\bea
A&=&\frac{ig_c^2}{2}(2\pi)^{p+1}\delta^{p+1}(p_1+p_2)\{\frac{1}{2s-1}
-\frac{(-2-t/2)}{2s}\cr
&&+\frac{\frac{1}{2!}(-2-t/2)(-3-t/2)}{2s+1}+\cdots\}\;\; ,\nonumber
\eea
Comparing it with the corresponding amplitude in the open string field theory
(\ref{cortach}), one finds that the tachonic pole is exactly the same. For
the massless pole we write $-2-t/2=p_1\inn N\inn p_2-s$. However, the term
proportional to $s$ in the massless pole gives contact term.
So we get, up to some contact terms, exact agreement in two cases.
Similarly, for the first massive pole one may rewrite it as
\bea
\frac{1}{2}(-2-\frac{t}{2})(-3-\frac{t}{2})&=&\frac{1}{2}
(p_1\inn N\inn p_2)^2-(p_1\inn N\inn p_2)(s+\frac{1}{2})
+\frac{1}{2}(s+\frac{1}{2})^2-\frac{1}{8}\; .
\nonumber
\eea
Again  terms that are proportional to $(2s+1)$ give contact term. 
After dropping these terms, one finds exactly the massive pole as 
(\ref{cortach}).

\subsection{Graviton amplitude}

The scattering amplitude of two massless closed string states from D-brane 
in the bosonic string theory is evaluated in \cite{CLR}. The result is
\bea
A&=&\frac{ig_c^2}{4}(2\pi)^{p+1}\delta^{p+1}(p_1+p_2)
\{e_1B(-t/2,1+2s)+e_2B(-t/2,2s)
\cr
&&-e_3B(1-t/2,2s)+e_4B(1-t/2,1+2s)+e_5B(-1-t/2,1+2s)\cr
&&+e_6B(1-t/2,-1+2s)+e_7B(-1-t/2,-1+2s)-e_8B(-t/2,-1+2s)\cr
&&-e_9B(2-t/2,-1+2s)+e_{10}B(3-t/2,-1+2s)\}\;\; ,
\label{SMAT}
\eea
where the kinematic factors $e_1,\cdots,e_{10}$ are\footnote{Note that 
our convention for matrix D,V,N  is different from those in \cite{CLR}, 
\ie $D_{\rm here}=-V_{\rm there},\; N_{\rm here}=D_{\rm there},\; 
V_{\rm here}=N_{\rm there}$.}
\bea
e_1&=&\frac{1}{2}p_2\inn\veps_1^T\inn\veps_2\inn p_1+
\frac{1}{2}p_2\inn\veps_1\inn\veps_2^T\inn p_1+p_2\inn\veps_1\inn 
D\inn\veps_2\inn p_2\cr
&&+p_2\inn\veps_1^T\inn\veps_2\inn D\inn p_2+p_2\inn\veps_1\inn
\veps_2^T\inn D\inn p_2+(1\leftrightarrow 2)\;\; ,\cr
e_2&=&-\tr(\veps_1\inn D)p_1\inn\veps_2\inn p_1+
(1\leftrightarrow 2)\;\; ,\cr
e_3&=&-p_1\inn D\inn\veps_1\inn D\inn\veps_2\inn D\inn p_2+
\frac{1}{2}p_1\inn D\inn\veps_1^T\inn\veps_2\inn D\inn p_2\cr
&&+\frac{1}{2}p_1\inn D\inn\veps_1\inn\veps_2^T\inn D\inn p_2-
\tr(\veps_1\inn D)
p_1\inn D\inn\veps_2\inn D\inn p_1 +(1\leftrightarrow 2)\;\; ,\cr
e_4&=&\tr(\veps_1\inn D\inn\veps_2\inn D)-p_1\inn D\inn\veps_2\inn D\inn\veps_1
\inn D\inn p_2-p_2\inn D\inn\veps_1\inn D\inn\veps_2\inn D\inn p_1
\;\; ,\cr
e_5&=&\tr(\veps_1\inn\veps_2^T)-p_1\inn\veps_2^T\inn\veps_1\inn p_2-p_2
\inn\veps_1\inn\veps_2^T\inn p_1\;\; ,\cr
e_6&=&\frac{1}{2}\tr(\veps_1\inn D)\tr(\veps_2\inn D)-
\tr(\veps_1\inn D)p_2\inn D\inn\veps_2\inn D\inn p_2\cr
&&+(p_2\inn\veps_1\inn p_2)(p_1\inn D\inn\veps_2\inn D\inn p_1)
+\frac{1}{2}(p_2\inn\veps_1\inn D\inn p_2)(p_1\inn\veps_2\inn D\inn p_1)
\cr
&&+(p_2\inn\veps_1\inn D\inn p_2)(p_1\inn D\inn\veps_2\inn p_1)+\frac{1}{2}
(p_2\inn D\inn \veps_1\inn p_2)(p_1\inn D\inn\veps_2\inn p_1)
+(1\leftrightarrow 2)\; ,\cr
e_7&=&(p_1\inn\veps_2\inn p_1)(p_2\inn\veps_1\inn p_2)\;\; ,\cr
e_8&=&-(p_2\inn\veps_1\inn p_2)(p_1\inn\veps_2\inn D\inn p_1+p_1\inn 
D\inn\veps_2\inn p_1)+(1\leftrightarrow 2)\;\; ,\cr
e_9&=&-(p_2\inn D\inn\veps_1\inn D\inn p_2)(p_1\inn\veps_2\inn D\inn p_1
+p_1\inn D\inn \veps_2\inn p_1)+(1\leftrightarrow 2)\;\; ,\cr
e_{10}&=&(p_1\inn D\inn\veps_2\inn D\inn p_1)(p_2\inn D\inn\veps_1\inn 
D\inn p_2)\;\; .\nonumber
\eea
Since in \cite{CLR} the authors were interested in the low energy contact 
terms of the amplitude from which 
the low energy effective action can be found, the 
above amplitude was expanded in a limit in which $s,t\longrightarrow 0$. 
On the other hand, since in our point of view we are not looking for the 
low energy expansion, instead, we expand the amplitude in the $s$-channel 
using the expansion (\ref{beta}) for the beta functions, that is
\bea
A&=&\frac{ig_c^2}{4}(2\pi)^{p+1}\delta^{p+1}(p_1+p_2)\{\frac{1}{2s-1}
(e_6+e_7-e_8-e_9+ e_{10})\cr
&&+\frac{1}{2s}[e_2-e_3+2e_7-e_8+e_9-2e_{10}+\frac{t}{2}
(e_6+e_7-e_8-e_9+e_{10})]\cr
&&+\frac{1}{2s+1}[e_1+e_2+e_4+e_5+3e_7-e_8+e_{10}+\frac{t^2}{8}
(e_6+e_7-e_8-e_9+ e_{10})\cr
&&+\frac{t}{2}(e_2-e_3+\frac{1}{2}e_6+\frac{5}{2}e_7
-\frac{3}{2}e_8+\frac{1}{2}
e_9-\frac{3}{2}e_{10})]+\cdots\}\;\; ,\nonumber
\eea
where dots represent the $s$-channel poles for higher massive modes.
Using the conservation of momentum on the world-volume of the D-brane, 
we recognize that the above $s$-channel poles, up to some contact terms, 
are exactly identical to the poles appear in the string field theory 
amplitude (\ref{agrav}). This ends our illustration of the equality of 
the scattering  amplitude of two closed string states from D-brane in 
the open string field theory and  bosonic string theory.
\section{Conclusion}
In this paper we studied the closed string scattering amplitude off a 
D-brane in 
the framework of the open string field theory. Using the CFT method we 
have been able to show, explicitly, that the gauge invariant observable 
in open string field theory introduced in \cite{{HN},{GRSW}}
reproduces the correct pole structure of the scattering amplitude in the 
bosonic string theory. 
We have checked this in the level truncation method up to level two. 
We note, however, that although the $s$-channel poles can be directly obtained 
from the level truncation computation, to see the $t$-channel poles we 
need to evaluate the scattering amplitude to all 
level in the open string field. 
This can be seen from the form of the beta function (\ref{beta}).
In fact each term in this expansion 
has a pole in $s$, but to see a pole in $t$ 
we need to take into account infinite terms in the summation. 
In principle one can compute the scattering amplitude to all level and then  
resume the expansion finding the exact expression for the $S$-matrix as 
(\ref{SMAT}) up to some contact terms. 
Therefore we conclude that the scattering amplitude of the 
closed string states in the open string field theory framework is the same 
as one in the bosonic string theory framework up to some contact terms. 
This contact terms might be related to a field redefinition.

As an other example of the scattering 
amplitude one can also compute the interaction of
one closed string and two open string fields. 
In the open string field theory, 
this corresponds to 
$\langle\; {\cal O}\;\Psi *\Psi *\Psi\;\rangle$.
The result should be the same as the corresponding amplitude in the bosonic
string theory up to some contact terms. We left the detail computation of this 
case for the future work.

Probably more interesting problem is to find the gauge invariant operators in 
the open supersting field theory. 
Although the superstring field theory action is
not as simple as the bosonic one, but since in the 
gauge invariant operator only
the string field and on-shell closed string 
are involved one might suspect that the
gauge invariant operator has the same form as the bosonic one. 
If this were that 
case one would proceed to compute the scattering 
amplitude in the superstring case.

Finally we note that there are two approaches 
in string field theory: (1) Open string
field theory which we have used throughout 
this paper and (2) boundary string field 
theory originally introduced in \cite{{WITTEN2},{SH}} 
(for recent discussions see \cite{MMK}).
It is believed that these two string field theories are equivalent. 
More precisely
there should be a field redefinition, though singular, which maps open string 
field theory to the boundary string field theory. Therefore an interesting 
question one might ask is what is the gauge invariant operators in the 
boundary string field theory?
This gauge invariant operator would have the same 
role as one in the open string
field theory, namely it should correspond to the 
on-shell closed string states \footnote{ The closed string in the BSFT has been
studied in \cite{SH1}.}.

{\large {\bf Acknowledgments}}

We would like to thank H. Arfaei and A. Ghodsi for 
discussions. M. A. would also like to
thank Institute for theoretical Physics (ITFA), Amsterdam, for hospitality.


\begin{thebibliography}{99}

\bibitem{KO}
K. Ohmori, ``A Review on Tachyon Condensation in Open String Field Theories'',
hep-th/0102085.

\bibitem{ABGKM}
I.Ya. Aref'eva, D.M. Belov, A.A. Giryavets, A.S. Koshelev and P.B. Medvedev,
``Non-commutative Field Theories and (Super)String Field Theories'',
hep-th/0111208. 

\bibitem{WITTEN}
E. Witten, `` Non-commutative Geometry and String Field Theory'', 
{\em Nucl Phys.}
{\bf B268} (1986) 253.

\bibitem{SEN}
A. Sen, `` Descent Relations Among Bosonic D-branes'',  {\em Int.J.Mod.Phys.}
{\bf  A14}  (1999) 4061; hep-th/9902105, and `` Universality of the 
Tachyon Potential'',  
{\em JHEP} {\bf 9912} (1999) 027;
hep-th/9911116.

\bibitem{FGST}
D. Z. Freedman, S. B. Gidding, J. A. Shapiro and C.B. Thorn, ``The nonplanar
one-loop amplitude in Witten's String field theory'',  
{\em Nucl. Phys.} {\bf 298}
(1988) 253.

\bibitem{TU}
H. Terao and S. Uehara, ``On the dilaton vertex in the 
Covariant Formulation of String'',
{\em Phys. Lett. } {\bf B188} (1987) 198.

\bibitem{ST1}
J. A. Shapiro  and C. B. Thorn, ``BRST invariant Transition between Closed and 
Open String'',
{\em Phys. Rev.} {\bf D36} (1987) 432, and ``Closed String-Open String 
Transition and Witten's String Field Theory'', {\em Phys. Lett.} 
{\bf B194} (1987) 43.

\bibitem{STO}
A. Strominger, ``Closed String in Open String Field Theory'',
 {\em Phys. Rev. Lett.} {\bf 16} (1987) 629.

\bibitem{ZW}
B. Zwiebach, ``Interpolation String Field Theories'', 
{\em Mod. Phys. Lett.} {\bf A7} (1992) 1079; hep-th/9202015.

\bibitem{HN}
A. Hashimoto and N. Itzhaki, ``Observable of String Field Theory'', 
hep-th/0111092.

\bibitem{GRSW}
 D. Gaiotto, L. Rastelli, A. Sen and B. Zwiebach, `` Ghost Structure and 
 Closed Strings in Vacuum String Field Theory'',  hep-th/0111129.

\bibitem{GM}
M. R. Garousi and R. C. Myers,  ``Superstring Scattering from D-Branes'',
 {\em Nucl.Phys.} {\bf  B475} (1996) 193; hep-th/9603194.

\bibitem{LPP}
A. LeClair, M.E. Peskin and C. R. Preitschopf, ``String Field Theory on the 
Conformal Plane (I). Kinematical Principles'', {\em Nucl. Phys.} 
{\bf B317} (1989) 411, and ``String Fields Theory on the 
Conformal Plane (II). Generalized Gluing'',  {\em Nucl. Phys.} 
{\bf B317} (1989) 464.

\bibitem{RZ}
 L. Rastelli, B. Zwiebach, ``Tachyon potentials, star products and 
 universality'',  {\em JHEP} {\bf 0109} (2001) 038; hep-th/0006240.

\bibitem{KT}
I. R. Klebanov and L.Thorlacius. ``The size of $p$-brane'', 
{\em Phys. Lett.} {\bf B371} (1996) 51; hep-th/9510200.

\bibitem{CLR}
S. Corley, D. Lowe and  S. Ramgoolam, ``Einstein-Hilbert action on the 
brane for the bulk graviton'', {\em JHEP} {\bf 0107} (2001) 030; 
hep-th/0106067.

\bibitem{WITTEN2}
E. Witten, ``On Background Independent Open-String Field Theory'', 
{\em Phys.Rev.}
{\bf D46} (1992) 5467; hep-th/9208027, and
``Some Computations in Background Independent 
Open-String Field Theory'', {\em Phys.Rev.} {\bf D47} (1993) 3405; 
hep-th/9210065.

\bibitem{SH}
S. L.  Shatashvili, ``Comment on the Background 
Independent Open String Theory'',
{\em Phys.Lett.} {\bf B311} (1993) 83; hep-th/9303143, and 
``On the Problems with Background Independence in String Theory'', 
hep-th/9311177.

\bibitem{MMK}
A. A. Gerasimov and S. L. Shatashvili, ``On Exact Tachyon Potential in 
Open String Field Theory'', {\em JHEP} {\bf 0010} (2000) 034; hep-th/0009103.

D. Kutasov, M. Marino and G. Moore, ``Some 
Exact Results on Tachyon Condensation 
in String Field Theory'', {\em JHEP} {\bf 0010} (2000) 045; hep-th/0009148, and
`` Remarks on Tachyon Condensation in 
Superstring Field Theory'', hep-th/0010108.

\bibitem{SH1}
A. A. Gerasimov and S. L. Shatashvili, `` String Higgs
Mechanisam and Fate of Open string'', hep-th/0011009, and
S. L. Shatashvili, ``On Field Theory of Open Strings,
Tachyon Condensation and closed Strings'',hep-th/0105076.

\end{thebibliography}
\end{document}